\documentclass{INTERSPEECH2023}
\usepackage{multirow}

\interspeechcameraready

\title{A Study on Prosodic Entrainment in Relation to Therapist Empathy in Counseling Conversation}
\name{Dehua Tao$^1$, Tan Lee$^1$, Harold Chui$^2$, Sarah Luk$^2$}
\address{
  $^1$ Department of Electronic Engineering \quad 
   $^2$ Department of Educational Psychology\\The Chinese University of Hong Kong}

\email{dhtao@link.cuhk.edu.hk, tanlee@ee.cuhk.edu.hk, \{haroldchui, sarah\_luk\}@cuhk.edu.hk}

\begin{document}

\maketitle
 
\begin{abstract}
Counseling is carried out as spoken conversation between a therapist and a client. The empathy level expressed by the therapist is considered an important index of the quality of counseling and often assessed by an observer or the client. This research investigates the entrainment of speech prosody in relation to subjectively rated empathy. Experimental results show that the entrainment of intensity is more influential to empathy observation than that of pitch or speech rate in client-therapist interaction. The observer and the client have different perceptions of therapist empathy with the same entrained phenomena in pitch and intensity. The client's intention to make adjustment on pitch variation and intensity of speech is considered an indicator of the client's perception of counseling quality.

\end{abstract}
\noindent\textbf{Index Terms}: counseling conversation, therapist empathy, prosodic entrainment, synchrony

\section{Introduction}
\label{sec:intro}

Counseling typically takes the form of spoken conversation between a therapist and a client. It aims to help the client to express thoughts and feelings, lower psychological distress, and make changes in life. In the field of psychotherapy, empathy is a natural human ability described as ``the therapist's sensitive ability and willingness to understand the client's thoughts, feelings, and struggles from the client's point of view" \cite{rogers1995way}. Therapist empathy is considered an essential indicator of counseling outcomes \cite{elliott2018therapist, moyers2013low, elliott2011empathy}. The empathy level expressed by the therapist over the course of a counseling conversation is often evaluated either by the observer (expressed empathy) or the client (received empathy). 

Entrainment is an important aspect of human-human communication, defined as the similarity of communicative behaviors between speakers \cite{wynn2022classifying}. This phenomenon correlates with the communicative success of interactive dialogue by achieving mutual comprehension and understanding \cite{pickering2006alignment, garrod2004conversation, pickering2004toward}, and building rapport and reducing social distance \cite{de2014investigating, miles2009rhythm, lakin2003using}. The relationship between entrainment and empathy has been studied attentively \cite{arizmendi2011linking, decety2004functional, preston2002empathy}. In particular, prosodic entrainment between therapist and client was found to be related to the therapist empathy \cite{schoenherr2021influence, xiao15b_interspeech, imel2014association, xiao13_interspeech}. In the present study, we analyze the prosodic entrainment in counseling conversations and its relation to the therapist empathy perceived by an observer or the client. As representative acoustic features of speech prosody, pitch, intensity and speech rate are analyzed.


Speech entrainment tends to occur as a dynamic phenomenon, i.e., the similarity of speech behaviors between interlocutors may increase or decrease over the course of a conversation \cite{wynn2022classifying, de2014investigating, levitan11_interspeech}. Considering such dynamicity, we propose to divide a conversation into overlapped sections based on speaker turns. A speaker turn refers to the time period during which only one person speaks. A conversation section is defined as a sequence of $N$ consecutive speaker turns. The synchrony of prosody and the degree of entrainment at the section level are measured. Here synchrony refers to the similarity in the variation of prosodic features between two speakers in terms of the direction and magnitude of change \cite{wynn2022classifying, levitan11_interspeech}. For example, as one speaker raises their pitch, the interlocutor does the same.

It is noted that the incidence of entrainment in client-therapist interaction varies with different prosodic cues. Therapists tend to adjust their speech intensity in accordance with that of the clients more often than to adjust pitch or speech rate. Toward the same speech behaviors of the client and therapist in counseling, the observer and the client have different perceptions about how they correlate with therapist empathy. In addition, experimental results suggest that observing how the client responds to the therapist can be used as a complementary cue to help assess the client's perception of counseling quality.

The remainder of the paper is organized as follows: Section~\ref{sec:dataset} describes the counseling conversations and empathy ratings used in this work. Section~\ref{sec:method} introduces the extraction of prosodic features and the measurement of prosodic entrainment. Section~\ref{sec:results} presents the correlations between entrainment and therapist empathy. Section~\ref{sec:disc} discusses the results, followed by conclusion in Section~\ref{sec:conc}.

\section{Counseling Speech Dataset}
\label{sec:dataset}

The counseling conversations analyzed in this study come from a speech dataset named CUEMPATHY \cite{tao2022cuempathy}. The audio recordings were collected during the counseling practicum for therapist trainees at a university in Hong Kong. Clients were adults who came to seek reduced-fee counseling services over a range of concerns related to stress, emotion, relationship, self-esteem, personal growth, and career. The dataset consists of $156$ counseling conversations from $39$ pairs of therapists and clients ($39$ distinct therapists and $39$ distinct clients), i.e., each pair contributes $4$ conversations. All therapists and clients are native Cantonese speakers. In a typical conversation, the therapist and the client take turn to speak. Each conversation is about $50$ minutes long. On average there are $316$ speaker turns in each conversation. The study was approved by the institutional review board, and informed consent was obtained from both the therapists and clients in written form.

Two measures of therapist empathy are analyzed in this work. They are the Therapist Empathy Scale (TES) \cite{decker2014development} given by the observer, and the Barrett-Lennard Relationship Inventory (BLRI) \cite{barrett2015relationship} given by the client. Another client-rated measure, Session Evaluation Scale (SES) \cite{hill2002development} is also included in our analysis. The SES reflects the overall effectiveness of counseling from the client's perspective. Table~\ref{tab:rating_info} provides information about the three ratings. For more details, please refer to \cite{tao2022cuempathy}.

\begin{table}[htb]
\caption{Description of three ratings for counseling sessions.}
\label{tab:rating_info}
\centering
\resizebox{1.0\linewidth}{!}{
\begin{tabular}{c|c|c|c|c}
Measure & Rater    & \begin{tabular}[c]{@{}c@{}}No. of\\items\end{tabular} & \begin{tabular}[c]{@{}c@{}}Point scale of\\each item\end{tabular}                                    & \begin{tabular}[c]{@{}c@{}}Range of\\total score\end{tabular} \\ \hline\hline
TES     & Observer & 9           & \begin{tabular}[c]{@{}c@{}}$1=\textit{not at all}$\\to $7=\textit{extremely}$\end{tabular}                                 & 9 to 63                                                         \\ \hline
BLRI    & Client   & 16          & \begin{tabular}[c]{@{}c@{}}$-3=\textit{strongly disagree}$\\to $+3=\textit{strongly agree}$\\ (0 is not used)\end{tabular} & -48 to +48                                                      \\ \hline
SES     & Client   & 5           & \begin{tabular}[c]{@{}c@{}}$1=\textit{strongly disagree}$\\to $5=\textit{strongly agree}$\end{tabular}                     & 5 to 25                                                         \\ 
\end{tabular}}
\end{table}

Clients were asked to give both BLRI and SES ratings after each counseling session. For TES, eight observer raters with counseling training at master level or above were recruited. As a reliability check, about $40\%$ ($62$ sessions) of the recorded sessions were rated by two observers. The intra-class coefficient (ICC) based on a mean-rating ($k = 2$), consistency, and two-way random effects model was $0.90$. According to the Cicchetti’s guidelines\footnote{The Cicchetti’s guidelines on inter-rater reliability: ICCs $< 0.40$ mean poor, $0.40 - 0.59$ mean fair, $0.60 - 0.74$ mean good, and $> 0.75$ mean excellent reliability.} \cite{cicchetti1994guidelines}, this value of ICC means excellent inter-rater reliability. 

The audio recording of a counseling conversation is divided into speaker-turn-based audio segments based on manually annotated speaker turns and speech transcriptions (in traditional Chinese characters). Turn-level speech-text alignment is performed as described in \cite{tao2022cuempathy}. Table~\ref{tab:data_info} gives the summary of $155$ counseling conversations used in the following experiments. One conversation was dropped because the BLRI and SES ratings were missing.


\begin{table}[htb]
\caption{Summary of speech content in $155$ counseling conversations. Average speech time per conversation (AvgST), average duration per turn (AvgDur), and average number of characters per turn (AvgChar) are given for each speaker.}
\label{tab:data_info}
\centering
\resizebox{1.0\linewidth}{!}{
\begin{tabular}{c|c|c|c}
Speaker       & AvgST (min) & AvgDur (sec) & AvgChar \\ \hline\hline
Therapist (T) & 15.04                                                                                & 6.38                                                                       & 26                                                                           \\ \hline
Client (C)    & 33.42                                                                                & 16.40                                                                      & 64                                                                           \\
\end{tabular}}
\end{table}

\section{Method}
\label{sec:method}

\subsection{Prosodic features}

Prosodic features are computed from the speech in each speaker turn. Pitch and intensity values are extracted on short-time frame basis using the openSMILE toolkit \cite{eyben2010opensmile} on each turn. The median, mean, and standard deviation of frame-level pitch over the turn are computed as the turn-level pitch parameters. The median and mean indicate the overall pitch level of the turn. The standard deviation is used to measure the degree of pitch variation over the turn. The same set of statistical measures is computed for frame-level intensity. The speech rate is measured in terms of the number of characters (syllables) spoken per second. The turn-level speech rate is computed by dividing the number of syllables contained in the turn by the time duration of the turn. The turn-level feature parameters are normalized on speaker basis by subtracting the mean of raw turn-wise parameters from that speaker in the conversation.

\subsection{Entrainment measurement}
\label{entrmeas}

To capture entrainment dynamics during the course of a counseling conversation, the sequence of speaker turns in the conversation is divided into overlapped sections. Each section contains $N$ consecutive turns. The step size between neighboring sections is $M$. It is assumed that the similarity between the speech behaviors of client and therapist is constant in a section, and it varies from one section to another. The local static entrainment is measured from adjacent turns in each section \cite{wynn2022classifying}. Since our focus is to observe how the therapist ($T$) responds to the client ($C$) during counseling, speaker order of the turn sequence in a section is denoted as $C, T, C, T, ...$. That is, the client's turn is followed by the therapist's turn. It is further assumed that (1) $N$ and $M$ are even numbers, and (2) each section begins with the client's turn (odd indices for the client and even for the therapist). Otherwise, the first and/or last turn of the conversation can be chopped to satisfy the assumptions. In the following paragraphs, we will explain how to analyze the synchrony of prosody and measure the degree of entrainment at section level.

\subsubsection{Synchrony of prosody}


Section-level synchrony and anti-synchrony are measured by the Spearman correlation. Anti-synchrony refers to the phenomenon that contrasts synchrony. For each section, the Spearman correlation coefficient between two sets of turn-level prosodic parameters, representing the client and the therapist respectively, is computed. A large value of positive correlation indicates strong synchrony as reflected by the prosodic features in the section. A large value of negative correlation reveals there exists a strong anti-synchronous relation between the observed features. Other values of correlation manifest that neither synchrony nor anti-synchrony is exhibited in the interaction between client and therapist. The threshold for significant positive or negative correlations is set to be \textpm$0.5$ at a significant level of $0.05$ in the experiments.

Borrowing the idea from \cite{de2014investigating}, the ratios of synchronous/anti-synchronous states are calculated to quantify the entrainment of client-therapist interaction during the conversation. The ratio of synchronous state is obtained by (1) merging neighboring synchronous sections; (2) counting the number of turn pairs (i.e., $C$ and $T$) in merged sections; (3) dividing the number of turn pairs in the synchronous state by the total number of turn pairs in the conversation. The ratio of anti-synchronous state is computed in a similar way. To analyze how empathy ratings are related to states of synchrony/anti-synchrony, the Pearson correlation between each of these two ratios and each type of empathy rating is calculated. Such analysis can reveal the level of synchrony with different prosodic features would have different effects on the empathy perceived by the observer or the client.

\subsubsection{Degree of entrainment}

The degree of entrainment in each section is measured as suggested in \cite{xiao15b_interspeech}, i.e., computing the averaged absolute difference of turn-level prosodic parameters between the client and the therapist. Let $x(i)$ and $x(i+1)$ denote the feature parameters of two consecutive speaker turns that belong to the client and the therapist, respectively. The averaged absolute difference $D_x$ of a section is defined by Eq. (\ref{eq1}) below. The mean and standard deviation of section-wise differences are computed over a counseling conversation. The relationship between the mean or standard deviation and each type of empathy rating across counseling conversations is analyzed by the Pearson correlation.

\begin{equation}
\begin{aligned}
  D_x = \frac{1}{N/2}\sum_{i=1}^{N/2} |x(2i) - x(2i - 1)|
  \label{eq1}
\end{aligned}
\end{equation}

\section{Results}
\label{sec:results}

\subsection{States of synchrony/anti-synchrony in counseling}
\label{subsec:syncres}

With the assumption that the local entrainment in a section is constant, the choice of section size may have an impact on the result of entrainment analysis. Thus, different section sizes, i.e., $N\in\{20,30,40,50\}$, with a fixed step size of $M=10$, are attempted in the experiments. The correlation results obtained from different combinations of $N$ and $M$ are reported according to the significant level. For example, the correlation coefficient between the synchronous state in pitch median and the TES score at $N=40$ is reported in Table~\ref{tab:ratio_corr} because its significant level is higher than at other values of $N$. This principle also applies to Section~\ref{subsec:degreeres}.

\begin{figure}[htb]
  \centering
  \includegraphics[width=0.8\linewidth]{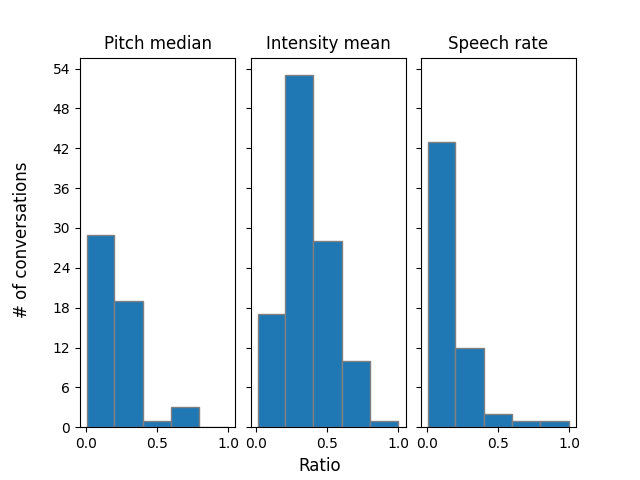}
  \caption{Ratio of synchronous state in pitch median, intensity mean, and speech rate.}
  \label{fig:sync_hist}
\end{figure}

\begin{figure}[htb]
  \centering
  \includegraphics[width=0.8\linewidth]{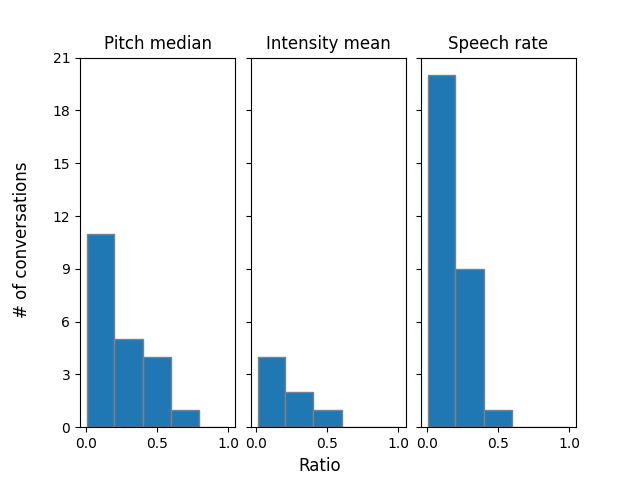}
  \caption{Ratio of anti-synchronous state in pitch median, intensity mean, and speech rate.}
  \label{fig:antisync_hist}
\end{figure}

Taking the pitch median, intensity mean, and speech rate at $N=40$ as examples, the distributions of ratios of synchronous and anti-synchronous states across all $155$ counseling conversations are depicted in Figure~\ref{fig:sync_hist} and \ref{fig:antisync_hist} respectively. It is observed that the synchronous state is more common than the anti-synchronous state in client-therapist interaction during counseling. In addition, the occurrence rate of synchronous state is different among different prosodic cues. For example, synchrony may occur more frequently in intensity than pitch or speech rate.


%
%

Table~\ref{tab:ratio_corr} shows the results of Pearson correlation analyses between ratios of synchronous/anti-synchronous states and three ratings (i.e., TES, BLRI, and SES) for each of the prosodic parameters. The ratio of synchronous state in pitch level is positively correlated with the observer-rated TES score while negatively correlated with the client-rated BLRI and SES scores. This indicates that when the client raises (or lowers) the pitch level, the observer tends to perceive it as an empathic behavior if the therapist also raises (or lowers) the pitch level. However, the client may feel that the therapist is not expressing empathy. Neither synchrony nor anti-synchrony in pitch variation shows a correlation with empathy ratings. In terms of the intensity level, the ratio of synchronous state is negatively correlated with all three ratings. Thus, as the client speaks progressively louder (or softer), it would be a positive sign of empathy if the therapist decreases (or increases) the volume. For intensity variation, the synchrony in client-therapist interaction may be a negative sign for observer-rated and client-rated empathy. In addition, it would be a non-empathic behavior from the observer's perspective (i.e., TES) if the therapist adjusts their speech rate in accordance with that of the client.

\begin{table}[htb]
\caption{Correlations between ratios of synchrony/anti-synchrony and three ratings. Significant level: $^*$ for $p<.1$, $^{**}$ for $p<.05$, and $^{***}$ for $p<.01$. ``N.S." refers to ``Not Significant".}
\label{tab:ratio_corr}
\centering
\resizebox{1.0\linewidth}{!}{
\begin{tabular}{c|ccc|ccc}
\multirow{2}{*}{Feature}                                   & \multicolumn{3}{c|}{Synchrony} & \multicolumn{3}{c}{Anti-synchrony} \\ \cline{2-7} 
                                                           & TES      & BLRI      & SES     & TES       & BLRI       & SES       \\ \hline\hline
\begin{tabular}[c]{@{}c@{}}Pitch\\ Median\end{tabular}     & $0.232^{***}$ & $-0.240^{***}$ & $-0.138^*$ & N.S.      & N.S.       & N.S.      \\ \hline
\begin{tabular}[c]{@{}c@{}}Pitch\\ Mean\end{tabular}       & N.S.     & $-0.198^{**}$  & N.S.    & N.S.      & N.S.       & N.S.      \\ \hline
\begin{tabular}[c]{@{}c@{}}Pitch\\ Std\end{tabular}        & N.S.     & N.S.      & N.S.    & N.S.      & N.S.       & N.S.      \\ \hline\hline
\begin{tabular}[c]{@{}c@{}}Intensity\\ Median\end{tabular} & N.S.     & $-0.159^{**}$  & $-0.133^*$ & $0.172^{**}$   & N.S.       & N.S.      \\ \hline
\begin{tabular}[c]{@{}c@{}}Intensity\\ Mean\end{tabular}   & $-0.197^{**}$ & N.S.      & N.S.    & $0.166^{**}$   & $0.133^*$     & N.S.      \\ \hline
\begin{tabular}[c]{@{}c@{}}Intensity\\ Std\end{tabular}    & $-0.185^{**}$ & N.S.      & N.S.    & N.S.      & $0.326^{***}$   & $0.251^{***}$  \\ \hline\hline
\begin{tabular}[c]{@{}c@{}}Speech\\ Rate\end{tabular}      & N.S.     & N.S.      & N.S.    & $-0.137^*$   & N.S.       & N.S.     
\end{tabular}
}
\end{table}

\subsection{Degree of entrainment in relation to empathy}
\label{subsec:degreeres}

\begin{figure}[t]
  \centering
  \includegraphics[width=0.8\linewidth]{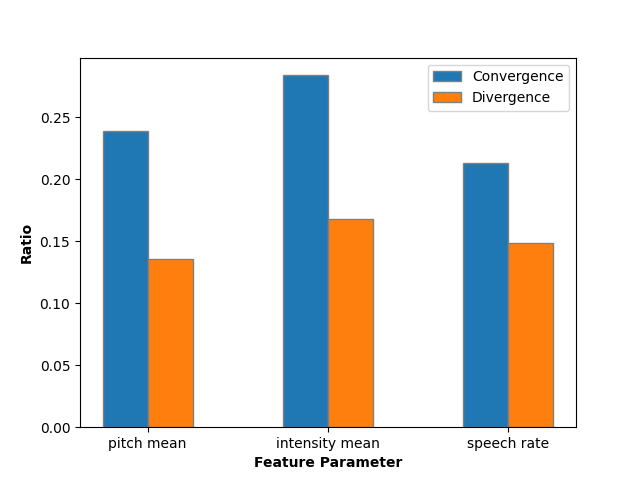}
  \caption{Ratios of conversations with section-wise differences showing the trend of convergence or divergence for different prosodic features.}
  \label{fig:sess_ratio}
\end{figure}

In order to investigate if section-wise averaged absolute differences increase or decrease continuously over the course of a counseling conversation, the Pearson correlation coefficient between section-wise differences and section time is computed. The significant level is set at $0.05$. Figure~\ref{fig:sess_ratio} illustrates the percentages of conversations that show significant positive (i.e., convergence) or negative (i.e., divergence) correlations for different prosodic features at $N=40$. The section-wise differences are found to be convergent or divergent in only $37.4\%$ ($58$ conversations), $45.2\%$ ($70$ conversations), and $36.1\%$ ($56$ conversations) of total $155$ conversations for pitch mean, intensity mean, and speech rate, respectively. This confirms that the prosodic entrainment often dynamically evolves in most counseling conversations.


The mean and standard deviation of section-wise differences are calculated to represent the degree and the fluctuation of entrainment over the course of a conversation, respectively. The relationship between each of the two statistics and each of the three empathy ratings is analyzed through the Pearson correlation. The significant correlation coefficients are reported in Table~\ref{tab:diff_corr}. Both the mean and standard deviation of section-wise differences in pitch parameters positively correlate with the SES score given by the client. This suggests that the lower degree and greater fluctuation in pitch entrainment in client-therapist interaction may be more helpful in improving the counseling quality from the client's perspective. However, the standard deviation of section-wise differences in pitch level is found to be negatively correlated with the TES score given by the observer. Thus it may be considered a positive sign of observer-rated empathy that the degree of entrainment in pitch level remains stationary throughout a conversation. In addition, it is noted that higher entrainment in intensity level between the client and the therapist is associated with higher observer-rated empathy but lower client-rated empathy. The degree of entrainment in intensity variation is negatively correlated with the client ratings. Similar results are found in the standard deviation of intensity differences. For speech rate, the greater fluctuation of entrainment is found to be associated with higher empathy from the observer's perspective.

\begin{table}[htb]
\caption{Correlations between mean and standard deviation of section-wise differences and three ratings. Significant level: $^*$ for $p<.1$, $^{**}$ for $p<.05$, and $^{***}$ for $p<.01$. ``N.S." refers to ``Not Significant".}
\label{tab:diff_corr}
\centering
\resizebox{1.0\linewidth}{!}{
\begin{tabular}{c|ccc|ccc}
\multirow{2}{*}{Feature}                                   & \multicolumn{3}{c|}{Mean}     & \multicolumn{3}{c}{Standard Deviation} \\ \cline{2-7} 
                                                           & TES     & BLRI     & SES      & TES          & BLRI       & SES        \\ \hline\hline
\begin{tabular}[c]{@{}c@{}}Pitch\\ Median\end{tabular}     & N.S.    & N.S.     & $0.174^{**}$  & $-0.195^{**}$     & N.S.       & $0.156^*$     \\ \hline
\begin{tabular}[c]{@{}c@{}}Pitch\\ Mean\end{tabular}       & N.S.    & N.S.     & $0.164^{**}$  & $-0.186^{**}$     & N.S.       & $0.215^{***}$     \\ \hline
\begin{tabular}[c]{@{}c@{}}Pitch\\ Std\end{tabular}        & N.S.    & N.S.     & $0.133^*$   & N.S.         & N.S.       & $0.178^{**}$    \\ \hline\hline
\begin{tabular}[c]{@{}c@{}}Intensity\\ Median\end{tabular} & $-0.154^*$ & $0.146^*$   & $0.168^{**}$  & $-0.235^{***}$    & N.S.       & N.S.       \\ \hline
\begin{tabular}[c]{@{}c@{}}Intensity\\ Mean\end{tabular}   & $-0.139^*$ & $0.181^{**}$  & $0.191^{**}$  & $-0.234^{***}$    & $0.203^{**}$    & $0.211^{***}$   \\ \hline
\begin{tabular}[c]{@{}c@{}}Intensity\\ Std\end{tabular}    & N.S.    & $0.228^{***}$ & $0.262^{***}$ & N.S.         & $0.218^{***}$   & $0.224^{***}$   \\ \hline\hline
\begin{tabular}[c]{@{}c@{}}Speech\\ Rate\end{tabular}      & N.S.    & N.S.     & N.S.     & $0.233^{***}$     & N.S.       & N.S.      
\end{tabular}
}
\end{table}

\section{Discussion}
\label{sec:disc}


Experimental results show that the observer and the client may have different perceptions of the same speech behaviors in relation to therapist empathy in counseling. For example, when the pitch levels of the client and the therapist change in a synchronous manner during a conversation, the observer tends to see this as a positive sign of empathy, while the client does not. The higher entrainment in intensity level between the client and the therapist is associated with higher observer-rated empathy but lower client-rated empathy. In addition, the correlation analyses among TES, BLRI, and SES scores show that there is no significant correlation between the observer-rated TES and the client-rated BLRI ($r=-0.07$) or SES ($r=-0.07$). However, a strong positive correlation exists between BLRI and SES ($r=0.72$). The discrepancy between observer and client ratings reveals that different raters may focus on different aspects of counseling when making their rating decisions. This issue has also been mentioned in previous studies \cite{elliott2011empathy, gurman1977patient}.

To investigate whether the way the client reacts to the therapist in terms of prosodic features is related to empathy ratings, prosodic synchrony analyses are conducted in the context of the turn order of $T, C, T, C, ...$ in a section, i.e., the therapist's turn is followed by the client's turn. There is no significant correlation between synchrony or anti-synchrony and observer-rated empathy. This suggests that the observer tends to focus on how the therapist responds to the client when evaluating the empathy level. However, synchrony in pitch variation is found to be positively correlated with client-rated BLRI and SES scores. This correlation is not found in client-therapist interaction. In addition, synchrony in intensity level is also negatively correlated with BLRI scores, consistent with the result reported in Table~\ref{tab:ratio_corr}. This indicates that observing how the client interacts with the therapist can also help assess the client's perceptions of counseling quality.

As mentioned in Section~\ref{subsec:syncres}, the results reported in Table~\ref{tab:ratio_corr} and \ref{tab:diff_corr} are obtained from different combinations of $N\in\{20,30,40,50\}$ and $M=10$. It has been found that the section size affects the correlation analysis of prosodic synchrony. For some feature parameters, significant correlations between states of synchrony or anti-synchrony and empathy ratings are found only for a specific value of $N$. However, the section size has little effect on measuring the degree of prosodic entrainment because the absolute difference is averaged over C-T turn pairs in a section as shown in Eq. (\ref{eq1}).

\section{Conclusion}
\label{sec:conc}

In this work, we investigate the prosodic entrainment between the client and the therapist and its relation to therapist empathy in the counseling conversation. Synchrony of prosody and the degree of entrainment are measured on turn pairs of client-therapist at the section level. It is observed that the occurrence rate of synchrony in client-therapist interaction is varied for different prosodic features. The synchronous state in intensity is more often exhibited than that in pitch or speech rate during counseling. Experimental results indicate that the observer and client may give opposite ratings when observing the same entrainment behaviors in the conversation, including synchrony in pitch level and high entrainment in intensity level. In addition, it is found that the way the client responds to the therapist in terms of pitch variation and intensity level also reflects the client's perception of counseling quality.

\section{Acknowledgements}
\label{sec:ack}

This research is partially supported by the Sustainable Research Fund of the Chinese University of Hong Kong (CUHK) and an ECS grant from the Hong Kong Research Grants Council (Ref.: 24604317).



\bibliographystyle{IEEEtran}
\bibliography{mybib}

\end{document}